\documentclass[conference]{IEEEtran}
\IEEEoverridecommandlockouts
\usepackage{cite}
\usepackage{listings}
\usepackage{xcolor}
\usepackage{stmaryrd}
\usepackage{tikz}
\usepackage[utf8]{inputenc}

\usepackage{orcidlink}

\usepackage{url}

\usepackage{caption}
\usepackage{amsmath,amssymb,amsfonts}
\usepackage{graphicx}
\usepackage{textcomp}
\usepackage{algorithm}
\usepackage{algpseudocode}
\usepackage{placeins}
\def\BibTeX{{\rm B\kern-.05em{\sc i\kern-.025em b}\kern-.08em
    T\kern-.1667em\lower.7ex\hbox{E}\kern-.125emX}}

\usepackage{microtype}
\usepackage{booktabs}

\newcommand{\contractsem}[1]{\llbracket #1\rrbracket}
\newcommand{\bracepipe}[1]{%
  \tikz[baseline=(x.base)]{
    \node[inner sep=0pt] (x) {$\left\{#1\right\}$};
    \draw[line width=0.3pt] ([xshift=-0.73ex,yshift=1.65ex]x.base) -- 
      ([xshift=-0.73ex,yshift=-0.55ex]x.base);
    \draw[line width=0.3pt] ([xshift=0.73ex,yshift=1.65ex]x.base) -- 
      ([xshift=0.73ex,yshift=-0.55ex]x.base);
  }%
}

\newcommand{\hwsem}[1]{\bracepipe{#1}}

\begin{document}

\title{Coverage-Guided Leakage Contract Fuzzing for Open-Source Processors}
\title{Coverage-Guided Pre-Silicon Fuzzing of\\ Open-Source Processors based on Leakage Contracts}

\author{\IEEEauthorblockN{Gideon Geier}
\IEEEauthorblockA{\textit{Saarland University} \\
\textit{Saarland Informatics Campus}\\
Saarbrücken, Germany\\
gideon.geier@gmail.com}
\and
\IEEEauthorblockN{Pariya Hajipour}
\IEEEauthorblockA{\textit{Department of Computer Engineering} \\
\textit{Sharif University of Technology}\\
Tehran, Iran \\
paria.hajipour@sharif.edu}
\and
\IEEEauthorblockN{Jan Reineke\orcidlink{0000-0002-3459-2214}}
\IEEEauthorblockA{\textit{Saarland University} \\
\textit{Saarland Informatics Campus}\\
Saarbrücken, Germany\\
reineke@cs.uni-saarland.de}
}

\maketitle

\begin{abstract}
Hardware-software leakage contracts have emerged as a formalism for specifying side-channel security guarantees of modern processors, yet verifying that a complex hardware design complies with its contract remains a major challenge. While verification provides strong guarantees, current verification approaches struggle to scale to industrial-sized designs. 
Conversely, prevalent hardware fuzzing approaches are designed to find functional correctness bugs, but are blind to information leaks like Spectre~\cite{kocher2020spectre}.

To bridge this gap, we introduce a novel and scalable approach: coverage-guided hardware-software contract fuzzing. Our methodology leverages a self-compositional framework to make information leakage directly observable as microarchitectural state divergence. The core of our contribution is a new, security-oriented coverage metric, Self-Composition Deviation (SCD), which guides the fuzzer to explore execution paths that violate the leakage contract. We implemented this approach and performed an extensive evaluation on two open-source RISC-V cores: the in-order Rocket Core and the complex out-of-order BOOM core. Our results demonstrate that coverage-guided strategies outperform unguided fuzzing and that increased microarchitectural coverage leads to a faster discovery of security vulnerabilities in the BOOM core.\looseness=-1
\end{abstract}


\section{Introduction}

Modern high-performance processors, with complex features like out-of-order, speculative execution, and deep memory hierarchies, have introduced a new class of critical security vulnerabilities. Attacks such as Spectre\cite{kocher2020spectre} have demonstrated that microarchitectural optimizations, previously considered implementation details, can be exploited to leak sensitive information, breaking fundamental security boundaries between software processes. 

In response, the research community has proposed hardware-software leakage contracts~\cite{DBLP:conf/sp/GuarnieriKRV21,wang2023specification} to formally capture the security guarantees a processor must provide, specifying exactly what information may leak. However, ensuring that a complex hardware design actually adheres to its contract is challenging.

On the one hand, \emph{contract verification}~\cite{wang2023specification,wang2025,DBLP:conf/asplos/TanYBM025,Hsiao21,DBLP:conf/micro/HsiaoNKMPFT24} %
can offer mathematical proof of a design's security. While rigorous, these methods still face intractable scalability issues and become prohibitively expensive when applied to industrial-scale cores. 

On the other hand, \emph{conventional hardware fuzzing}~\cite{hur2021difuzzrtl,DBLP:conf/nanoarch/FuAJG21,DBLP:conf/uss/SoltCR24}, while more scalable, addresses a different property: functional correctness.
Such fuzzers employ differential testing to find functional correctness bugs.
Their objective is to ensure the final architectural state (i.e., registers and memory) matches a trusted reference model. 
This approach does not address side-channel vulnerabilities as these often do not constitute functional bugs; the processor produces the correct final result while leaking information through attacker-observable side effects that a standard architectural reference model is blind to.\looseness=-1

Recently, a new class of contract-aware fuzzers~\cite{oleksenko2022revizor, oleksenko2023hide,hofmann2023speculation,oleksenko2025enter,Nemati2020a,buiras2021micro} focusing on speculative leakage has emerged that directly targets side-channel vulnerabilities.
Their goal is similar to ours: to identify contract violations, i.e., to find a program with a pair of inputs, such that the two executions are indistinguishable at the level of the contract, but distinguishable to an attacker at the hardware level.
However, existing contract-aware fuzzers are applied to commercial processors at the post-silicon stage. 
Lacking a model of the internal behavior of these processors, they treat the hardware as a black box and approximate the abilities of attackers via established side-channel attacks such as Prime+Probe.

There is also a growing body of pre-silicon hardware fuzzers~\cite{rajapaksha2023sigfuzz, hur2022specdoctor} that target side-channel vulnerabilities.
However, these approaches are focused on relatively narrow classes of attacks and are not contract aware.

In software fuzzing, \emph{coverage feedback} has been used to increase fuzzing efficiency, e.g. in AFL~\cite{afl}.
However, without a model of the internal behavior of commercial processors, such an approach cannot be taken at the post-silicon stage.
In this work, addressing open-source processors at the pre-silicon stage, we introduce coverage guidance into hardware-software contract fuzzing. 

We leverage a self-compositional testing framework, where two instances of the same processor are simulated in lockstep with different secret data, to make information leakage directly observable as a divergence in execution. The core innovation of our approach is that we guide this fuzzing process using a new, security-oriented coverage metric called \textit{Self-Composition Deviation (SCD)}. This metric tracks divergences in the microarchitectural state between the two processor instances, allowing our fuzzer to prioritize test cases that explore execution paths that are more likely to cause security-critical contract violations.

\pagebreak
Our work makes the following contributions:
\begin{enumerate}
    \item We introduce coverage-guided feedback to contract-aware fuzzing by defining a novel microarchitectural coverage metric, Self-Composition Deviation (SCD), specifically designed to measure state-space exploration relevant to information leakage.

    \item We design and implement a fuzzing framework, demonstrating that coverage-guided feedback improves fuzzing efficiency. We propose and evaluate multiple strategies, showing that a \textit{Weighted Feedback} approach, which prioritizes test cases covering rare and unique execution paths, is more effective at exploring the design space than unguided methods.

    \item We conduct an extensive experimental evaluation on two widely-used, open-source RISC-V cores: the simple in-order Rocket Core and the complex out-of-order BOOM core. Our results confirm that our approach is effective: Coverage-guided fuzzing achieves superior coverage, and increased coverage leads to faster detection of contract violations in the vulnerable BOOM core.
\end{enumerate}

\section{Background}

\subsection{Hardware-Software Leakage Contracts}
\label{sec:2a}

The aim of hardware-software leakage contracts is to capture a processor's information leakage due to microarchitectural side channels at the software level.
Capturing leakage at the software level enables programmers and compilers to check whether the execution of a given program on a particular processor leaks any secret information or not, \emph{without} having to explicitly consider the processor's microarchitectural implementation.

We associate the ``software level'' with the processor's instruction set architecture (ISA).
The ISA defines the architectural state, consisting of the state of the memory and the architecture's registers, and its dynamics, i.e., how the architectural state evolves upon the execution of instructions.
Popular ISAs including ARM and RISC-V have been formalized in domain-specific languages like ASL~\cite{DBLP:conf/fmcad/Reid16} and Sail~\cite{armstrong2019isa}.
Such formal specifications may thus serve as the basis for hardware-software leakage contracts.
The ISA is the target language of compilers and compiler-correctness proofs relate the semantics of the source program to the semantics of the target program in terms of the ISA.
Leakage contracts at ISA level thus serve as a natural target for constant-time-preserving compilers such as Jasmin~\cite{DBLP:journals/pacmpl/OlmosBBGL25}.

To express leakage at ISA level we augment an ISA semantics by associating potential leakage with each transition, i.e., each execution of an instruction.
For example, in order to capture leakage via a data cache, a contract may expose the memory address accessed upon every load and store instruction.
To express operand-dependent timing of division instructions, a contract could expose the instruction's operand values.\looseness=-1

In this work, we will focus on two contracts from~\cite{DBLP:conf/sp/GuarnieriKRV21}, detailed later:
The \emph{seq-ct} contract, which encodes common constant-time programming assumptions, and the \emph{seq-arch} contract, which captures guarantees provided by secure-speculation mechanisms.\looseness=-1

Testing or verifying these contracts involves checking the hardware's actual implementation against the specified guarantees.
To facilitate such approaches, we next define contracts more formally.

\newcommand{\arch}{\textsc{Arch}}
\newcommand{\muarch}{\textsc{µArch}}
\newcommand{\obs}{\textsc{Obs}}

A \textit{hardware-software leakage contract} augments the semantics of an Instruction Set Architecture (ISA) by introducing contract observations to the transitions between architectural states. Let $\sigma = \langle m, a\rangle \in \arch$ be an \textit{architectural state}, which consists of the memory $m$ and a register assignment $a$. %
The architectural semantics is captured by a deterministic binary relation $\to\ \subseteq \arch \times \arch$, mapping an architectural state to its successor state during execution.
As we assume that the program is immutable during its execution, it is not part of the architectural state in our formalization.
However, it determines the transitions that may be taken.
Thus, technically, each program $p$ defines a particular relation $\to_p$.
We sometimes omit the subscript $p$ where this does not introduce ambiguity.

Contracts augment architectural semantics by associating transitions with contract observations.
This is captured by a ternary relation $\rightharpoonup_p \subseteq \arch \times \obs \times \arch$, such that $\sigma \overset{l}{\rightharpoonup_p} \sigma'$ if $\sigma \to_p \sigma'$ and $l$ is the contract observation associated with the transition from $\sigma$ to $\sigma'$.

Then, a \textit{contract trace} is a finite sequence of contract observations obtained along a run of the architectural semantics.
For simplicity, we only consider terminating executions here.
Let $p$ be a program and $\sigma_0 \in \arch$ be an initial state.
Then, we denote by $\contractsem{p}(\sigma_0)$, $p$'s contract trace from $\sigma_0$:
\begin{equation} \label{eq:contract_trace}
\contractsem{p}(\sigma_0) = l_1, l_2, \dots, l_n ~\text{where }
\sigma_0 \overset{l_1}{\rightharpoonup_p} \sigma_1 
\overset{l_2}{\rightharpoonup_p}  \dots 
 \overset{l_n}{\rightharpoonup_p} \sigma_{n}
\end{equation}

\newcommand{\attacker}{{\cal{A}}}
\newcommand{\atkobs}{\textsc{AtkObs}}

Two executions that produce the same contract trace should be indistinguishable to an attacker.
In order to formalize contract satisfaction, we need to capture the abilities of an attacker.
To this end, we introduce a microarchitectural semantics $\Rightarrow_p\ \subseteq (\arch \times \muarch) \times (\arch \times \muarch)$ that models the hardware behavior at cycle granularity upon executing program~$p$.
It operates on pairs of architectural states $\sigma \in \arch$ and microarchitectural states $\mu \in \muarch$.
Based on the microarchitectural semantics, we obtain \emph{hardware traces} as follows:
\begin{multline}
    \hwsem{p}(\sigma_0, \mu_0) = (\sigma_0, \mu_0), (\sigma_1, \mu_1), \dots, (\sigma_n, \mu_n)\\ ~\text{where } (\sigma_0, \mu_0) \Rightarrow_p (\sigma_1, \mu_1) \Rightarrow_p \dots \Rightarrow_p (\sigma_{n}, \mu_n)
\end{multline}

We model the attacker via a function $\attacker$ on traces generated by the microarchitectural semantics, i.e., $\attacker : (\arch \times \muarch)^* \rightarrow \atkobs$, where $\atkobs$ is the set of possible observations.
Different attacker models have been proposed in the past, including models that allow the attacker to see the cache state at some level of abstraction~\cite{DBLP:journals/tissec/DoychevKMR15}.
In this work we focus on a simple, yet powerful attacker model that reveals the number of cycles executed: $\atkobs(\tau) = |\tau|$.

Then, an \emph{attacker trace} is the result of the application of the attacker model to a hardware trace:
\begin{equation}
    \hwsem{p}_{\attacker}(\sigma_0, \mu_0) = \attacker(\hwsem{p}(\sigma_0, \mu_0))
\end{equation}

\subsubsection*{Contract Satisfaction}

For a hardware implementation to satisfy a contract under attacker~\attacker, its attacker traces must not reveal more information than the corresponding contract traces.\looseness=-1

Formally, contract satisfaction is given by:
\begin{equation} \label{eq:contract_satisfaction}
\begin{aligned}
\forall p, \forall \mu, \forall \sigma_0, \sigma_1 : \; & \contractsem{p}(\sigma_0) = \contractsem{p}(\sigma_1) \\
& \implies \hwsem{p}_{\attacker}(\sigma_0, \mu) = \hwsem{p}_{\attacker}(\sigma_1, \mu)
\end{aligned}
\end{equation}
This condition ensures that if two initial architectural states $\sigma_0$ and $\sigma_1$ produce identical contract traces, the attacker traces for the program must also be identical.

\subsection{Hardware Fuzzing}

Hardware fuzzing extends the principles of software fuzzing to pre-silicon hardware verification, aiming to uncover design bugs by exploring behavioral differences under varying inputs. Instead of searching for crashes or hangs, as in traditional software fuzzing, the focus is on detecting mismatches in hardware execution—such as differences in timing or register state—when the same program is executed with different data inputs. Our approach leverages RTL-level simulation to perform greybox fuzzing, allowing for efficient and automated testing.

Central to our method is the use of executable hardware-software contracts, built by extending the Sail model~\cite{armstrong2019isa} of the RISC-V ISA. These contracts define expected program behavior, enabling the detection of violations through a self-composed simulation: the same hardware design is executed in parallel with two different data sections. After each clock cycle, register states are compared, and any divergence or asymmetric termination signals a potential contract violation. We evaluate this approach on two RISC-V cores from the Chipyard\cite{chipyard} framework—Rocket and BOOM—and show that it can systematically identify information leaks via timing.

\subsection{Open-source RISC-V Cores and Tools}
In our work, we focus on two open-source RISC-V cores, both of which are part of the Chipyard framework:
\begin{itemize}
    \item \emph{Rocket} is a five-stage, in-order processor that implements the RV64GC instruction set. Developed as part of the Berkeley Rocket Chip Generator, it is designed for efficiency and simplicity. The core supports virtual memory and privilege levels, making it suitable for running full operating systems such as Linux. Due to its open-source nature and modular design, Rocket Core is a good target for evaluating hardware-software contracts and security properties.

    \item \emph{BOOM} (Berkeley Out-of-Order Machine) is a ten-stage out-of-order processor that also implements RV64GC. Unlike Rocket Core, which executes instructions in-order, BOOM leverages register renaming, out-of-order execution, and speculative execution to improve performance. Its out-of-order execution model introduces complexities that make it a compelling subject for our fuzzing framework, allowing us to analyze side-channel leakage and contract compliance in a high-performance RISC-V core.
\end{itemize}

\pagebreak
We also use the following tools and frameworks:
\begin{itemize}
    \item To facilitate hardware instrumentation and modification, we utilize \emph{FIRRTL} (Flexible Intermediate Representation for RTL), an intermediate representation designed for digital circuit designs\cite{Li:EECS-2016-9}. 
    FIRRTL serves as an abstraction layer for Chisel, a hardware description language embedded in Scala. In our work, we leverage FIRRTL to instrument processor designs before simulation, enabling observability enhancements that help detect violations of hardware-software contracts. 

    \item For formal ISA modeling, we employ \emph{Sail}, a domain-specific language (DSL) for specifying and simulating instruction set architectures. Sail allows us to model the RISC-V~\cite{armstrong2019isa} ISA and extend it with hardware-software leakage contracts. %

    \item For hardware simulation, we use \emph{Verilator}, an open-source Verilog simulator that converts Verilog/SystemVerilog into optimized C++ models. %
    By simulating RISC-V cores with Verilator, we can rapidly execute fuzzing campaigns while monitoring for violations.

    \item To orchestrate test execution and facilitate interaction with simulated hardware, we use \emph{cocotb} (Coroutine-Based Cosimulation Testbench), a Python-based framework for writing testbenches for Verilog/SystemVerilog designs. cocotb allows us to interface with Verilator, generate input stimuli, and analyze execution traces without needing Verilog-based testbenches. In our fuzzing framework, cocotb acts as the bridge between our fuzzing infrastructure and the simulated RISC-V cores.
\end{itemize}

We build upon \emph{DifuzzRTL}\cite{hur2021difuzzrtl} as the foundation for our fuzzing framework because it is the only open-source hardware fuzzer that provides memory simulation, which is essential for realistic execution within our testing environment. Since both Rocket and BOOM cores interact with the rest of the SoC via the TileLink Cached (TL-C) bus, a proper memory model is required to maintain execution coherence. DifuzzRTL already includes a Python-based TL-C protocol implementation, allowing seamless integration with Chipyard cores. Additionally, it provides a mutational fuzzer for RISC-V programs along with the necessary infrastructure to compile and execute them on the DUT, making it a well-suited starting point for our work. Built on the cocotb environment, DifuzzRTL is also compatible with Verilator, aligning well with our existing simulation framework. These features make it the most practical and extensible option for our fuzzing needs.

\section{Coverage-guided Contract Fuzzing}
To systematically find test cases that trigger contract-violating executions, a fuzzer must explore a processor's vast state space efficiently. A purely random or unguided approach is fundamentally limited, as it explores blindly without learning from past executions, often wasting cycles on redundant paths. To overcome this, our methodology is \emph{coverage-guided}, transforming the search from a random process into a feedback-driven exploration. The core principle is to monitor which microarchitectural states are exercised by a test case and use this coverage feedback to systematically prioritize new test cases that are more likely to uncover unique behaviors and, consequently, security violations.

\subsection{Fuzzing Architecture}

\begin{figure}[t]
    \centering
    \includegraphics[width=0.49\textwidth]{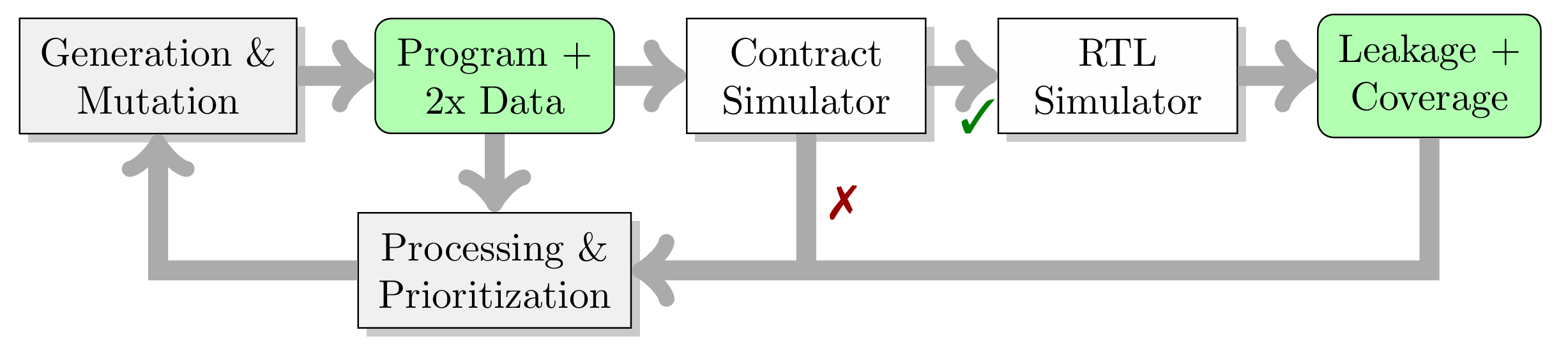}
    \caption{Fuzzing pipeline}
    \label{fig:pipeline}
\end{figure}

Our fuzzing framework follows a structured pipeline (Figure~\ref{fig:pipeline}) consisting of five key components: 1. generation and mutation, 2. contract simulation, 3. RTL simulation, 4. leakage and coverage analysis, and 5. processing/prioritization. 

\begin{enumerate}
    \item \emph{Generation and Mutation:}
    The fuzzing process begins with the generation and mutation stage, where programs are created and modified to test the processor. Each generated program consists of instruction sequences that exercise different execution behaviors within the processor. These programs are paired with two sets of data inputs, allowing us to compare execution variations across different inputs. The mutation engine modifies instructions, operands, memory accesses, and control flow structures to increase test coverage and expose vulnerabilities.\looseness=-1
    \item \emph{Contract Simulation:}
    Once the program and input data have been generated, they are passed to the contract simulator based on Sail. The contract simulator generates contract traces for the program on both data sections. If the traces differ, the runs are contract-distinguishable and we do not simulate it, but report back that the input was distinguishable. The contract-indistinguishable test cases are simulated and checked for contract violations.
    \item \emph{RTL Simulation:}
    The indistinguishable test cases are then executed in the RTL simulator, which models the actual microarchitecture of the processor. 
    We use Verilator for this step, as it provides cycle-accurate execution and enables high-performance simulation of the RISC-V cores. The RTL simulator captures timing behavior, memory access patterns, and architectural state transitions, allowing us to detect behaviors that may not be visible at the ISA level.
    Specifically, we implement the attacker model here, which is used in the following phase.\looseness=-1
    \item \emph{Leakage and Coverage Analysis:}
    After execution, the simulation results are passed into the leakage and coverage analysis component.
    Based on the attacker model, the pair of inputs is either attacker-distinguishable or not. If it is, a leak is found, as the pair violates contract satisfaction.
    Additionally, coverage metrics are collected to determine whether the generated test cases are effectively exploring new parts of the processor’s execution space. Programs that maximize coverage and expose new behaviors are flagged as high-priority for further mutation.
    \item \emph{Processing and Prioritization:}
    Finally, the processing and prioritization component refines the test-case set by determining which test cases should be mutated further and reintroduced into the fuzzing cycle. 
    Programs that led to new or rare coverage are given higher priority, based on the assumption that these are more likely to uncover leakage.
\end{enumerate}

\subsection{Hardware-Software Contracts in Sail}

To formally define the expected security guarantees of RISC-V microarchitectures, we implement hardware-software leakage contracts within the Sail framework. 
As defined earlier, these contracts define permitted information leakage at the ISA level.

For notational clarity, we write \texttt{$\sigma$, inst} to denote an architectural state \texttt{$\sigma$} in which \texttt{inst} is the next instruction to be executed. This abstracts away the explicit program representation, as \texttt{inst} is fully determined by the state and the program~$p$ to be executed.

Contract observations are sets of \texttt{(label, value)} pairs. 
Each \texttt{label} encodes the category or source of the leakage (e.g., \texttt{pc} or \texttt{lAddr}), while the corresponding \texttt{value} represents the concrete data exposed during the transition. 
This structure supports expressing multiple leakage events per instruction step. 
The set of observations may be empty indicating that no observable leakage occurs during that transition.

\subsubsection*{Sequential Leakage Models} For this work, we focus on two sequential contracts~\cite{DBLP:conf/sp/GuarnieriKRV21}, \emph{seq-ct} and \emph{seq-arch}, which express leakage in terms of the sequence of instructions executed non-speculatively, as defined in Section~\ref{sec:2a}.

\subsubsection*{Seq-ct - Constant-Time Execution Contract}
The seq-ct contract corresponds to common constant-time programming assumptions, allowing two types of leakage:
\begin{itemize}
    \item Program counter (PC): The program's control flow leaks by revealing the program counter. This captures leakage via the instruction cache.
    \item Memory addresses of load and store operations. This captures leakage via the data cache.
\end{itemize}

We define the seq-ct contract on top of the architectural semantics, expressed by the relation $\rightarrow_p$ for load and store instructions as follows:
\[
\scalebox{1.2}{$
\frac{
    \sigma,\, i\ \texttt{rd}\ \texttt{rs1}\ \texttt{imm} \rightarrow_p \sigma' 
    \quad 
    l = \left\{ \left( \mathit{lAddr},\, a(\texttt{rs1}) + \texttt{imm} \right),\, \left( {pc},\, a(\mathbf{pc}) \right) \right\}
}{
    \sigma,\, i\ \text{rd}\ \text{rs1}\ \text{imm} 
    \overset{l}\rightharpoonup_{p, \mathit{seq\text{-}ct}} 
    \sigma'
}
$}
\]
\hspace{33mm}\scalebox{0.9}{$
\text{for } i \in \{ \texttt{lb},\, \texttt{lh},\, \texttt{lw},\, \texttt{ld},\, \texttt{lbu},\, \texttt{lhu},\, \texttt{lwu} \}
$}

Here $\sigma = \langle m, a \rangle$ and thus $a$ refers to the state of the registers.

For store instructions:
\[
\scalebox{1.2}{$
\frac{
    \sigma,\, i\ \texttt{rs1}\ \texttt{rs2}\ \texttt{imm} \rightarrow_p \sigma'
    \quad
    l = \left\{ \left( \mathit{sAddr},\, a(\texttt{rs1}) + \texttt{imm} \right),\, \left( {pc},\, a(\mathbf{pc}) \right) \right\}
}{
    \sigma,\, i\ \texttt{rs1}\ \texttt{rs2}\ \texttt{imm} 
    \overset{l}\rightharpoonup_{p, \mathit{seq\text{-}ct}} 
    \sigma'
}
$}
\]
\hspace{55mm}
\scalebox{0.9}{$
\text{for } i \in \{ \texttt{sb},\, \texttt{sh},\, \texttt{sw},\, \texttt{sd} \}
$}

For all other instructions, only the PC is leaked:
\[
\scalebox{1.2}{$
\frac{
    \sigma \rightarrow_p \sigma' \quad l = \left\{ ( {pc},\, a(\mathbf{pc}) ) \right\}
}{
    \sigma \overset{l}\rightharpoonup_{p, \mathit{seq\text{-}ct}} \sigma'
}
$}
\scalebox{0.9}{$
\quad \text{otherwise}
$}
\]

To strengthen the guarantees of seq-ct, we introduce a stricter variant, seq-ct-b, which additionally leaks branch outcomes. This modification explicitly reveals whether a conditional branch was taken, allowing for more refined security verification.

\subsubsection*{Seq-ct-b - Extended Sequential Contract}
Seq-ct-b is a stricter variant of Seq-ct that additionally leaks the outcome of branch instructions:
\[
\scalebox{1.2}{$
\frac{
    \sigma,\, i\ \texttt{rs1}\ \texttt{rs2}\ \texttt{imm} \rightarrow_p \sigma'
    \quad
    l = \left\{ \left( \mathit{taken},\, a(\texttt{rs1}) \otimes a(\texttt{rs2}) \right) \right\}
}{
    \sigma,\, i\ \texttt{rs1}\ \texttt{rs2}\ \texttt{imm} 
    \overset{l}\rightharpoonup_{p, \mathit{seq\text{-}ct\text{-}b}} 
    \sigma'
}
$}
\]
\hspace{28mm}\scalebox{0.9}{$
\text{for } i \in \{ \texttt{beq},\, \texttt{bne},\, \texttt{blt},\, \texttt{bge},\, \texttt{bltu},\, \texttt{bgeu} \}
$}

\hspace{35mm}\scalebox{0.9}{$
\text{with } \otimes \in \{ =,\, \neq,\, <,\, \geq,\, <_u,\, \geq_u \}
$}

This additional leakage allows to distinguish executions that follow the same sequence of instructions even though the outcome of branch instructions was different. This is possible in corner cases where a taken branch leads to the same instruction as an untaken branch.

\subsubsection*{Seq-arch}

Seq-arch extends seq-ct, by additionally exposing the values of loads::
\[
\scalebox{1.2}{$
\frac{
    \sigma,\, i\ \texttt{rd}\ \texttt{rs1}\ \texttt{imm} \rightarrow_p \sigma'
    \quad
    l = \left\{ \left( \mathit{lValue},\, a'(\texttt{rd}) \right),\, \left( \mathit{lAddr},\, \ldots \right) \right\}
}{
    \sigma,\, i\ \texttt{rd}\ \texttt{rs1}\ \texttt{imm} 
    \overset{l}\rightharpoonup_{p,\mathit{seq\text{-}arch}} 
    \sigma'
}
$}
\]
\hspace{30mm}
\scalebox{0.9}{$
\text{for } i \in \{ \texttt{lb},\, \texttt{lh},\, \texttt{lw},\, \texttt{ld},\, \texttt{lbu},\, \texttt{lhu},\, \texttt{lwu} \}
$}

Here $a'$ refers to the value of the registers in the successor state $\sigma' = \langle m', a' \rangle$.

As a consequence, the results of all computations in registers are exposed to the attacker as we assume that they have full knowledge of the program.
This contract has been shown to be useful in the context of sandboxing untrusted code~\cite{DBLP:conf/sp/GuarnieriKRV21}.

\subsection{Coverage Generation}
\label{sec:covgen}

In order to facilitate efficient contract verification, we introduce a novel coverage metric, 
Self-Composition Deviation (SCD) coverage, which is specifically designed to focus on timing variations in microarchitectural states. 
This section formalizes the coverage model, presents its hardware instrumentation, and describes the software implementation 
for tracking and utilizing coverage feedback.

Traditional coverage metrics in hardware fuzzing focus on maximizing state space exploration. However, when fuzzing hardware-software leakage contracts, the primary goal is to detect deviations in the microarchitectural states in the two executions, which may eventually manifest in attacker distinguishability and thus contract violations. 
To this end, we introduce \emph{Self-Composition Deviation (SCD) coverage}, which measures differences between two concurrently executing 
instances of the same design under different inputs.

Formally, let $s_A$ and $s_B$ denote the states of two instances of the processor core, where both execute the same program with different data inputs. 
The state space divergence is captured by a deviation function:
\begin{equation}
    \Delta(s_A, s_B) = \{ r \mid r \in \text{Registers}, s_A[r] \neq s_B[r] \}
\end{equation}

SCD coverage tracks the evolution of $\Delta(s_A, s_B)$ across execution cycles.
To obtain the value of the deviation function in every cycle, the self-composition is instrumented to output a deviation bit vector, where each bit corresponds to a single register in the design, capturing $s_A[r] \neq s_B[r]$.

The instrumentation logic is implemented as a FIRRTL transformation, systematically duplicating modules and appending comparison logic for all registers.

The SCD coverage output of the simulated circuit is processed in software to guide fuzzing. 
Instead of tracking complete traces of deviation vectors, which would be prohibitively expensive in storage and processing, we employ a hashing mechanism:
\begin{equation}\label{eq:hash}
    \text{cov}[\text{hash}(\Delta(s_A, s_B)) \oplus \text{hash}(\Delta(s_A, s_B)_{t-1}) \gg 1] := 1
\end{equation}
At the start of the execution the \text{cov} vector of 2 MiB is initialized to 0 at every entry. 
Then, in each cycle, the current deviation vector ($\Delta(s_A, s_B)$) and the previous deviation vector ($\Delta(s_A, s_B)_{t-1}$) are hashed and combined as described in Equation~\ref{eq:hash} and the corresponding bit in the \text{cov} vector is set.\looseness=-1

In other words, a rolling hash of the deviation vector is maintained across execution cycles, reducing the required storage footprint while preserving coverage information. 
The implementation employs the SHAKE128 hashing algorithm.

The coverage vector obtained at the end of the execution of a test case $tc$ is used by the fuzzer to determine the priority of the test case. 
We denote $tc$'s coverage vector by $\text{cov}_{tc}$.
At any point in time, we also determine the \emph{cumulative coverage} of the set of test cases~$TC$ explored so far:
\begin{equation}\label{eq:cumulativecoverage}
    \forall i: \text{cumulative-cov}[i] = \bigvee_{tc \in TC} \text{cov}_{tc}[i]
\end{equation}
In Section~\ref{sec:prioritization}, we explore multiple strategies of prioritizing a test case based on its coverage and the cumulative coverage.

\section{Fuzzing Pipeline Implementation and\\ Design Decisions}

The fuzzing pipeline integrates various components to enable the identification of contract violations in processor cores. It follows an iterative loop where new test cases are generated using a mutator. Each iteration consists of the following steps:
\begin{enumerate}
    \item Generate a new test case.
    \item Compile and execute the test case using the contract simulator.
    \item If contract distinguishability is observed, store the test case and proceed to the next iteration.
    \item If not distinguishable, execute the RTL simulation to detect any violations.
    \item Update the fuzzer’s coverage data and prioritize new test cases accordingly.
\end{enumerate}

Pseudocode of the fuzzing loop is given below:

\begin{lstlisting}[language=Python, caption={Fuzzing Loop}]
def fuzz(num_iter):
  cum_coverage = bitarray(repeat(0,2 ** 24))
  for i in range(0, num_iter):
    # Test-case generation
    (program, (data_a, data_b)) = mutator.get()
    (hsc_input, rtl_input) = compile(program, data_a, data_b)

    # Contract-distinguishability check
    ret = run_contract_check(hsc_input)
    mutator.update_data_seed_energy(program.get_seed(), ret)
    if ret == CONTR_DIST:
      save_mismatch(i)
      continue

    # RTL simulation
    (ret, cov) = run_rtl_sim(rtl_input)
    if ret == LEAK:
      save_leak(i)

    # Check for new coverage, 
    #    if found add program to corpus
    new_coverage = ~cum_coverage & cov
    if new_coverage.any():
      write_cov_log()
      program.save()
      mutator.add_corpus(program)
      cum_coverage = cum_coverage | cov
\end{lstlisting} %

\subsection{Test-Case Generation}

The test-case generation process creates a program and a pair of inputs per iteration to simulate both sides of a self-composed system, allowing parallel execution with different data sections since the program under test is assumed to be public knowledge. Test-case generation involves detailed steps:

\subsubsection{Program Generation}

In this section, we describe our adaptation of the DifuzzRTL program generation method for our specific use case. The modifications made to DifuzzRTL's approach were relatively minor, yet significant for our testing goals. The main changes involved adding an argument to restrict the instruction set used during generation and removing support for interrupts and virtual memory.

The general process of program generation remains straightforward. The instruction generator randomly selects an opcode and constructs an \texttt{instruction word}, which is a sequence of up to four instructions. Each word contains one main instruction, placed last in the sequence, with preceding instructions inserted to ensure the correct execution context for the main instruction. For example, if the main instruction is a load, it must be preceded by instructions that ensure the load instruction will execute meaningfully.

Initially, instruction words contain placeholder arguments such as registers, immediates, and symbols. These placeholders are later populated with concrete values. DifuzzRTL's population process deviates from purely random generation in two critical ways. Firstly, jumps and branches are restricted to move only forward within the program, ensuring guaranteed termination by preventing loops. This is implemented by labeling each word and only allowing forward jumps and branches, as observed in generated examples.

Secondly, DifuzzRTL employs a specific probability-based approach in choosing registers and immediates. There is a 20\% chance that a previously used register or immediate will be reused, creating intentional data dependencies across multiple instructions. This is intended to mimic realistic program behaviors, where multiple instructions operate on shared data.

After generating the instruction words, a data section is created separately, which we discuss in the following section. The resulting generated program body is then embedded into a standard template, modeled after the RISC-V unit test suite~\cite{riscv-tests}.  This template includes essential skeleton code that ensures valid termination and initializes the registers with the first 31 values from the data section. 

Subsequently, we compile these programs into the Executable and Linkable Format (ELF) using gcc for contract checking. For the simulation runs, we convert these ELF files into hexadecimal dumps using elf2hex, making them compatible with our hardware simulation environment.

\subsubsection{Data Section Generation}
\label{sec:dgeneration}
For the generation of the data sections, we explored multiple approaches and evaluated their effectiveness. Each data section consists of 3072 bytes, meaning multiple cache lines can be loaded during execution. %
Corresponding data sections are generated alongside each program to facilitate testing. The following describes the three different methods we used to create these data sections.

The first approach, referred to as ``fully random," involves generating 64-bit words at random and assembling them into a data section. This method results in a very low probability of having identical data at the same memory address across two different data sections, ensuring maximum randomness in the test cases.

The second approach, ``fully random for seq-arch," was specifically designed for testing with seq-arch. Since the testing template loads the first 31 words from the data section, the likelihood of differences arising in the fully random approach is high. To address this, we modified the method to ensure that the first 31 words are identical between data sections while keeping the rest random.

A third approach, ``50-50, random-equal," balances randomness with structured data similarities. With a probability of 0.5, a data section is generated using the fully random method. In other cases, the first 31 registers are enforced to be equal, as in the second approach. Additionally, for all remaining words in the data section, each word has a 50\% probability of being identical to its counterpart from the other data section. This approach ensures a controlled mix of randomness and determinism, allowing for better coverage in fuzzing experiments.

\subsubsection{Mutation}
The fuzzer generates programs until it fills one-tenth of its corpus size. Once this threshold is reached, new random programs are generated with a probability of 0.1, while all other cases involve mutating existing corpus entries. DifuzzRTL employs two types of mutation: \emph{Mutate} and \emph{Merge}, each occurring with an equal probability of 0.45. Below, we describe these mutation processes in detail and present an approach for handling data sections within the context of contract verification. 

Mutate: For the mutation process, a seed program is selected uniformly at random. from the corpus. Each instruction word within the selected program is then processed with the following probabilities: it is retained with a probability of 0.5, deleted with a probability of 0.25, or a new word is inserted after it with a probability of 0.25. Any new words introduced are populated accordingly, while retained words remain unchanged.

Merge: In the merging process, two programs, \(p_1\) and \(p_2\), are selected uniformly at random. The data section from \(p_1\) is retained. A random index \(i\) is selected from the range \(0\) to \(\min(p_1, p_2)\). A temporary program is then constructed by combining a prefix of \(p_1\) up to index \(i-1\), a slice of \(p_2\) from index \(i\) excluding the last five words, and the final five words from \(p_1\). This newly created temporary program is subsequently mutated as per the mutation process described.

To manage data section storage, DifuzzRTL employs a Least Recently Used (LRU) strategy. When a data section is newly created or reused in a mutation, it is moved to the top of the LRU queue. If the storage reaches capacity and a new data section is needed, the least recently used data section at the bottom of the queue is discarded.

Some data sections are expected to be tightly coupled to their corresponding program logic. For example, if a new load instruction is introduced via mutation under the seq-arch contract, the test case is unlikely to pass contract verification if most words in its data section differ significantly from their counterparts. To address this, we introduce a fallback mechanism that discards ineffective data sections.

Our replacement mechanism operates as follows: If the number of contract passes (\(n_{\text{pass}}\)) minus the number of contract failures (\(n_{\text{fail}}\)) for a given data section is less than -10, we regenerate the data section. The values \(n_{\text{pass}}\) and \(n_{\text{fail}}\) are updated after every contract check using the \texttt{update\_data\_seed\_energy} function. This ensures that ineffective data sections are removed while preserving those that contribute positively to contract compliance.

\subsection{Contract Checking}
In case of Rocket and BOOM, DifuzzRTL uses the Spike ISA simulator as the golden reference model to verify the correctness of the simulated execution. In our case, we redefine the contract as the reference model, replacing the Spike simulator with an interface to the Sail contract simulator. This simulator generates a contract trace for both ELF files, and we check these trace files for equality to determine whether the given test case exhibits contract (in)distinguishability.

\subsection{RTL Simulator}
The DifuzzRTL RTL simulator executes a continuous Verilator simulation of the core, resetting the design before executing a new program. To achieve a proper reset, DifuzzRTL %
introduces a \textit{metaReset} to the design that clears all registers.

Early experiments revealed that this reset mechanism was insufficient for our application, leading to ``leaks" that were not reproducible. 
The likely cause was that memories such as caches and buffers were invalidated but not zeroed, influencing execution results.

Another crucial adaptation was modifying the simulator to support self-composition and defining what qualifies as a ``leak." These topics are discussed in the following sections.

\subsubsection{Adapting to the Self-Composition}
The RTL simulator connects cocotb Python code with the Verilator simulation. It emulates the TL-C protocol and memory behavior. Since we now work with two data sections, we require two TL-C channels and two memory backends. Most of the existing code was duplicated to accommodate both sides of the self-composition.

\subsubsection{Attacker Model}
Both cores in the self-composition share the same clock and reset, ensuring synchronization. This setup allows us to observe any execution differences immediately during simulation. 
Various attacker models could be implemented here; we chose one where the attacker can observe only program start and termination from an OS perspective.

More specifically, the RISC-V model main loop, checks whether the program terminated after executing the next instruction. This is done via the Host/Target Interface (HTIF), which is a protocol to connect a host to a simulation target. We only use it to perform termination ‘syscalls’. For this, the program under execution writes a \texttt{1} in the memory at the \texttt{tohost} address. 
We detect termination by repeatedly polling this memory region via TL-C requests. However, due to protocol execution delays, minor timing differences (a few clock cycles) may go unnoticed.

We focus on detectable leakage that can be exploited in real attacks. If minor timing variations can be amplified through repeated execution patterns, the fuzzer should uncover such patterns. %
Our setup allows future extensions to analyze additional attacker models, such as detecting deviations on the program counter (PC) or instruction retirements at finer granularity.

\subsubsection{Ensuring Reproducibility}
\label{sec:reproducability}
The reset mechanism initially caused non-reproducibility of detected leaks upon execution on a freshly reset microarchitecture. 
To resolve this, we restructured the fuzzing setup. The simulator is now a standalone cocotb and Verilator application, while the fuzzer is written in Python, invoking external processes to collect coverage and leakage feedback. 

Launching a fresh Verilator processor and performing additional I/O operations for each test case increases runtime.
As a result, simulating a single test case now takes about 1.5 seconds on the smaller RocketCore---significantly longer than the previous execution time of a few hundred milliseconds. 
To mitigate this slowdown, we parallelized the simulation to prevent the fuzzer from idling while waiting for results.

Our new fuzzing loop consists of two internal cycles. The first loop generates test cases, checks for contract distinguishability, and, if a test case is contract-indistinguishable, adds the simulation task to a \texttt{ProcessPoolExecutor}. After accumulating 32 tasks, the system transitions to the second loop, retrieving results and updating the mutator with new coverage information. Once all results are processed, control returns to the first loop. By leveraging multi-core processors, this approach significantly improves fuzzing efficiency, especially for complex cores like BOOM, where individual simulations take longer.

\subsection{Prioritization}
\label{sec:prioritization}
A potentially effective approach to improving a fuzzer's performance is to prioritize test cases based on their coverage vector. 
After calculating the coverage vector for each test case, test cases that contribute more new and unique coverage may be given higher priority, as they are more likely to explore previously untested parts.

We consider the following strategies to evaluate the effect of coverage feedback:
\begin{enumerate}
\item \emph{Pass Feedback}:\\All contract-indistinguishable test cases are added to the corpus, regardless of whether they increase coverage.\\
    We select test cases from the corpus for mutation uniformly at random.

\item \emph{New Coverage Feedback}:\\ Only test cases that introduce new coverage are added to the corpus. 
We track flipped bits in the coverage bit vector, and if a test case flips at least one previously unseen bit, it is considered to have generated new coverage.\\
    We select test cases from the corpus for mutation uniformly at random.

\item \emph{Weighted Feedback}:\\ As in the previous case, only test cases that introduce new coverage are added to the corpus.\\
Rather than selecting test cases from the corpus uniformly at random, we employ the following prioritization:
\begin{itemize}
    \item Each coverage element is assigned a weight:
    \\
    Let \( C(tc) := \{ c \mid \text{cov}_{tc}[c] = 1\} \) be the set of coverage elements that test case \( tc \in \text{Corpus} \) contributes. Let \( n_c := |\{tc \in \text{Corpus} \mid \text{cov}_{tc}[c] = 1\}| \) be the number of test cases in the corpus that cover the coverage element~\( c \). The weight \( W(c) \) of each coverage element \( c \) is defined as:
    \begin{equation}
W(c) = \frac{1}{n_c}     
    \end{equation}

    \item Each test case is assigned a score:
    \\
    The score for each test case \( tc \), denoted \( \text{Score}(tc) \), is calculated as the sum of the weights of the coverage elements it contributes:

\begin{equation}
\text{Score}(tc) = \sum_{c \in C(tc)} W(c)    \label{eq:first}
\end{equation}

\item Prioritization:\\In the mutation and merge phase, each test case is picked with a probability, proportional to its weight, i.e., with the following probability:
    \begin{equation}
        {p_{tc}} = \frac{Score_{tc}}{\sum_{t \in \text{Corpus}}Score_t}
    \end{equation}

\end{itemize}
    This strategy ensures that test cases contributing rare or unique coverage elements are given more importance in the fuzzing process.

    \item \emph{Pass Feedback (100)}: \\
    In all of the above strategies, the corpus has a size of 1000 test cases.
    If adding a new seed would exceed this limit, the oldest test case is removed from the corpus. \\
    Reducing the corpus size may be beneficial in experiments where violations are expected to be found more quickly, i.e., with fewer test cases, as a smaller corpus fills up faster, leading to more frequent removal of older seeds and prioritization of newer test cases.
    Thus, we additionally consider a variant of \emph{Pass Feedback} in which the corpus size is reduced to 100.
\end{enumerate}

\section{Experimental Evaluation}

The purpose of our experiments is to answer the following questions:
\begin{itemize}
    \item Does coverage feedback increase cumulative coverage?
    \item Does increasing cumulative coverage lead to faster detection of the first contract violation?
\end{itemize}
To answer these, we conducted a series of experiments on the Rocket and BOOM RISC-V cores using four distinct prioritization strategies.

\subsection{Experimental Setup}
All experiments were performed using the base RV64I instruction set. This choice was made to ensure a rigorous evaluation of the prioritization strategies. The RV64IM extension introduces timing-dependent multiplication units that cause violations to be found almost immediately. By excluding these trivial leaks, we force the fuzzer to explore more subtle execution paths, providing a clearer measure of each strategy's true effectiveness.

The experiments were designed around two distinct core-contract pairings, each serving a specific purpose:
\begin{itemize}
    \item \emph{Rocket Core with seq-ct-b:} This pairing serves as a ``hard target" benchmark. Based on prior results, the in-order Rocket core is likely to satisfy the seq-ct-b contract. This configuration allows us to evaluate how effectively each strategy can explore the state space of a seemingly secure design without the immediate feedback of finding leaks.
    \item \emph{BOOM Core with seq-arch:} This pairing acts as a ``bug-finding" benchmark. The complex, out-of-order BOOM core is known to violate sequential contracts due to speculative execution. Testing against the restrictive seq-arch contract provides a challenging scenario to measure how quickly each strategy can discover these known-to-exist, non-trivial vulnerabilities.
\end{itemize}

Each experimental configuration was tested over five runs of 10,000 iterations each to ensure statistical validity. We evaluated the four prioritization strategies described in Section \ref{sec:prioritization}: Pass Feedback (with a corpus size of 1000 and 100), New Coverage Feedback, and Weighted Feedback.

\subsection{Increasing Cumulative Coverage}
\label{sec:cumulative}
To answer our first question, we measured the total cumulative coverage discovered by each strategy.
By total cumulative coverage we mean the number of elements in the final cumulative coverage, i.e., $|\{i \mid \text{cumulative-cov}[i] = 1\}|.$
The results unequivocally show that coverage-guided feedback, particularly the weighted approach, yields greater coverage than approaches without such feedback.

\subsubsection{Rocket Core Analysis (seq-ct-b)}
The data from the Rocket core experiments (Tables \ref{tab:rocket_coverage}, \ref{tab:rocket_corpus}, and \ref{tab:rocket_pvalue}) demonstrates a clear performance hierarchy.

\begin{table}[h!]
    \renewcommand{\arraystretch}{1.2}
    \centering
    \fontsize{7}{8.4}\selectfont
    \caption{Total cumulative coverage achieved (Rocket Core, seq-ct-b),\\ medians across runs are indicated in bold}
    \label{tab:rocket_coverage}
    \begin{tabular}{lrrrrr}
    \toprule
    \textbf{Strategy} & \textbf{Run 1} & \textbf{Run 2} & \textbf{Run 3} & \textbf{Run 4} & \textbf{Run 5} \\
    \midrule
    Pass Feedback & 28,214 & 26,487 & 26,736 & 25,659 & \textbf{26,587} \\
    Pass Feedback (100) & 17,975 & 16,439 & 18,672 & \textbf{17,911} & 16,567 \\
    New Coverage     & 46,407 & 45,850 & 45,993 & 46,831 & \textbf{46,256} \\
    Weighted Feedback & 53,730 & 56,550 & 54,324 & 55,396 & \textbf{54,325} \\
    \bottomrule
\end{tabular}
\end{table}

\begin{table}[h!]
\renewcommand{\arraystretch}{1.2}
    \centering
    \fontsize{7}{8.4}\selectfont
\caption{Number of test cases that increased the cumulative coverage (Rocket Core, seq-ct-b)}
\label{tab:rocket_corpus}
\begin{tabular}{lrrrrr}
\toprule
\textbf{Strategy} & \textbf{Run 1} & \textbf{Run 2} & \textbf{Run 3} & \textbf{Run 4} & \textbf{Run 5} \\
\midrule
Pass Feedback & 2,704 & 2,526 & 2,662 & 2,456 & \textbf{2,616} \\
Pass Feedback (100) & 1,576 & 1,625 & 1,531 & 1,374 & \textbf{1,546} \\
New Coverage     & 5,218 & 5,163 & \textbf{5,182} & 5,173 & 5,272 \\
Weighted Feedback & 5,568 & 5,908 & 5,540 & \textbf{5,589} & 5,703 \\
\bottomrule
\end{tabular}
\end{table}

The Weighted Feedback strategy is the most effective, achieving a median coverage of 54,325, nearly double that of the Pass Feedback approach. 

The Mann-Whitney U test is a non-parametric test to determine the stochastic ranking of two distributions A and B. The test is non parametric in that it makes no assumption
about the underlying distributions unlike, e.g., the t-test.
The Mann-Whitney U test (Table~\ref{tab:rocket_pvalue}) confirms that the performance differences between the strategies are statistically significant ($p < 0.05$). 
This superior performance is achieved by prioritizing test cases that cover rare execution paths, which in turn allows the fuzzer to identify more unique, coverage-increasing test cases to add to its corpus. Table~\ref{tab:rocket_corpus} illustrates this efficiency directly: the Weighted and New Coverage strategies added a median of 5,589 and 5,182 valuable test cases, respectively—more than double the 2,616 test cases added by the standard Pass feedback strategy. 

\begin{table}[h!]
\renewcommand{\arraystretch}{1.2}
    \centering
    \fontsize{7}{8.4}\selectfont
\caption{Mann-Whitney U test p-values for coverage (Rocket Core)}
\label{tab:rocket_pvalue}
\begin{tabular}{lrrr}
\toprule
\textbf{Strategy} & \hspace{-3mm}\textbf{Pass Feedback (100)} & \textbf{Pass Feedback } & \textbf{New Coverage} \\
\midrule
Pass Feedback       & 0.008 & & \\
New Coverage        & 0.008 & 0.008 & \\
Weighted Feedback   & 0.008 & 0.008 & 0.008 \\
\bottomrule
\end{tabular}
\end{table}

\subsubsection{BOOM Core Analysis (seq-arch)}
The same trend holds for the more complex out-of-order BOOM core. The \emph{Weighted Feedback} strategy again achieved the highest median coverage, and the statistical tests confirm its superiority over all other methods (Table \ref{tab:boom_pvalue}).

\begin{table}[h!]
\renewcommand{\arraystretch}{1.2}
    \centering
    \fontsize{6}{7.2}\selectfont
\caption{Total cumulative coverage achieved (BOOM Core, seq-arch)}
\label{tab:boom_coverage}
\begin{tabular}{lrrrrr}
\toprule
\textbf{Strategy} & \textbf{Run 1} & \textbf{Run 2} & \textbf{Run 3} & \textbf{Run 4} & \textbf{Run 5} \\
\midrule
Pass Feedback & 1,891,266 & 1,889,579 & 1,875,854 & \textbf{1,883,416} & 1,849,282 \\
Pass Feedback (100)  & \textbf{1,133,674} & 1,223,106 & 1,219,373 & 1,108,384 & 1,040,014 \\
New Coverage        & 1,906,725 & 1,848,400 & 1,852,767 & 1,905,778 & \textbf{1,881,362} \\
Weighted Feedback   & 2,083,805 & 2,024,513 & 2,066,326 & 2,043,683 & \textbf{2,046,181} \\
\bottomrule
\end{tabular}
\end{table}

\begin{table}[h!]
\renewcommand{\arraystretch}{1.2}
    \centering
    \fontsize{7}{8.4}\selectfont
\caption{Mann-Whitney U Test p-values for coverage (BOOM Core)}
\label{tab:boom_pvalue}
\begin{tabular}{lrrr}
\toprule
\textbf{Strategy} & \hspace{-3mm}\textbf{Pass Feedback (100)} & \textbf{Pass Feedback} & \textbf{New Coverage} \\
\midrule
Pass Feedback       & 0.008 & & \\
New Coverage        & 0.008 & 0.421 &  \\
Weighted Feedback   & 0.008 & 0.008 & 0.008 \\
\bottomrule
\end{tabular}
\end{table}

\subsubsection{Conclusion on Coverage}
To effectively increase cumulative coverage, a guided prioritization strategy is essential. The most powerful method is \emph{Weighted Feedback}, as it actively incentivizes the exploration of unique and rarely-exercised execution paths, leading to a statistically significant increase in total discovered coverage.

\subsection{Impact of Coverage on Leak Detection}

To evaluate the impact of coverage on leak detection, we also conducted extensive experiments on the Rocket core. We tested it against both the seq-ct-b contract and the even less restrictive seq-ct contract, running each experiment with all feedback strategies for up to one million iterations. Despite this comprehensive fuzzing campaign, we were unable to find any leakage beyond the ones permitted by the contracts for the Rocket core. %
This result, while demonstrating the robustness of the in-order core against these specific contracts, means that our analysis of leak detection performance must focus on the BOOM core experiments, where contract violations were consistently detected.

The data from the BOOM core reveal an inverse correlation: strategies that achieve higher coverage find leaks faster.

\begin{table}[h!]
\renewcommand{\arraystretch}{1.2}
    \centering
    \fontsize{7}{8.4}\selectfont
\caption{Number of test cases to find first leak with medians across runs in bold}
\label{tab:boom_leak_five}
\begin{tabular}{lrrrrr}
\toprule
\textbf{Strategy} & \textbf{Run 1} & \textbf{Run 2} & \textbf{Run 3} & \textbf{Run 4} & \textbf{Run 5} \\
\midrule
Pass Feedback  & \textbf{452} & 63 & 1,880 & 1,158 & 105 \\
Pass Feedback (100)  & \textbf{1,155} & 903 & 589 & 1,944 & 1,490 \\
New Coverage Feedback        & \textbf{452} & 63 & 3,178 & 1,158 & 105 \\
{Weighted Feedback}   & \textbf{194} & 63 & 913 & 1,398 & 105 \\
\bottomrule
\end{tabular}
\end{table}

Table~\ref{tab:boom_leak_five} shows the number of test cases to find the first leak across five runs.
The \emph{Weighted Feedback} strategy, which achieved the highest coverage, also found the first leak the fastest, requiring a median of 194 test cases. 
Conversely, the \emph{Pass Feedback (100)} strategy, which had the lowest coverage, was the slowest, with a median of 1,155 test cases. 

It is noticeable in Table~\ref{tab:boom_leak_five} that for some runs, different strategies find the first leak under the same number of test cases. 
This is because we used the same seeds for different strategies to ensure reproducibility of the experiments and the feedback only starts to take effect once 10\% of the corpus are filled. 
As a result, the same inputs are generated at the beginning and if a leak is found early on this happens after the same number of test cases have been explored.

\begin{table}[h!]
\renewcommand{\arraystretch}{1.2}
    \centering
    \fontsize{7}{8.4}\selectfont
\caption{Mean and median of number of test cases to find first leak across 100 runs}
\label{tab:boom_leak_hundred}
\begin{tabular}{lrr}
\toprule
\textbf{Strategy} & \textbf{Mean} & \textbf{Median} \\
\midrule
Pass Feedback           & 704.19 & 338.5  \\
Pass Feedback (100)     & 1586.87 & 1077.5  \\
New Coverage Feedback   & 730.61 & 326  \\
{Weighted Feedback}     & 483.66 & 279  \\
\bottomrule
\end{tabular}
\end{table}

\begin{table}[h!]
\renewcommand{\arraystretch}{1.2}
    \centering
    \fontsize{7}{8.4}\selectfont
\caption{Mann-Whitney U Test p-values for test cases to first leak with 100 runs (BOOM Core)}
\label{tab:boom_leak_pvalue}
\begin{tabular}{lrrr}
\toprule
\textbf{Strategy} & \hspace{-5mm}\textbf{Pass Feedback (100)} & \textbf{Pass Feedback } & \textbf{New Coverage} \\
\midrule
 Pass Feedback       & 0.149 & & \\
New Coverage        & 0.016 & 0.749 & \\
Weighted Feedback   & 0.003 & 0.187 & 0.220 \\
\bottomrule
\end{tabular}
\end{table}

To increase the statistical power of the experiment, we performed 100 runs under each strategy. 
We report the mean and median of the number of test cases to find the first leak in Table~\ref{tab:boom_leak_hundred}.
In terms of both statistics, the \emph{Weighted Feedback} strategy outperforms the other strategies.
This is also confirmed by the Mann-Whitney U Test p-values in Table \ref{tab:boom_leak_pvalue}, which show that the performance of the weighted strategy is significantly better than the worst-performing strategy, \emph{Pass Feedback (100)} ($p = 0.003$).

\section{Related Work}

We now discuss related work around hardware-software contracts and hardware fuzzing.

\subsection{Hardware-Software Contracts}
Hardware-software leakage contracts were introduced by Guarnieri et al.~\cite{DBLP:conf/sp/GuarnieriKRV21} to capture the security guarantees provided by secure speculation mechanisms.
On the software side, similar models had previously been used to capture the ``constant-time programming'' discipline~\cite{bernstein_aes_attack,DBLP:conf/ccs/GarciaBY16,DBLP:conf/icisc/MolnarPSW05,AlmeidaBBDE16} and to reason about software-security on processors implementing speculative execution~\cite{DBLP:journals/pacmpl/OlmosBBGL25,DBLP:conf/sp/GuarnieriKMRS20}.

Recent work has work has focussed on verifying~\cite{wang2023specification,DBLP:journals/iacr/BloemGGHMP22,DBLP:conf/asplos/TanYBM025} and synthesizing~\cite{wang2025,Mohr24,DBLP:conf/micro/HsiaoNKMPFT24} leakage contracts for open-source \mbox{RISC-V} processors, while we apply fuzzing to find contract violations.
A recent contract synthesis approach~\cite{wang2025} employs bounded verification to find contract violations to rule out candidate contracts.
It may be interesting to explore whether fuzzing is more efficient in detecting violations, speeding up the synthesis process.

\subsection{Post-Silicon Hardware Fuzzing}
Hardware fuzzing can be classified into pre- and post-silicon approaches.
Post-silicon approaches interact directly with actual silicon chips, while pre-silicon approaches apply to formal models of hardware, like simulators or RTL designs.
For commercial processors, academic researchers are often forced to resort to post-silicon fuzzing due to the public unavailability of adequate models.

The post-silicon fuzzing work most related to our work is contract-aware fuzzing based on Revizor~\cite{oleksenko2022revizor}.
As in our work, the expected leakage is expressed via leakage contracts.
While we model an attacker's abilities via an attacker model that extracts attacker observations from microarchitectural states, Revizor approximates an attacker's abilities by implementing known attacks, such as Prime+Probe.
Revizor is shown to detect several known as well as unknown vulnerabilities of Intel and AMD processors, including Spectre, MDS, and LVI.
Subsequent work~\cite{oleksenko2023hide} significantly improved fuzzing efficiency by (a) filtering test cases that do not trigger microarchitecturally-visible speculation, and by (b) generating test cases that are by construction contract-indistinguishable.
This allowed a more extensive fuzzing campaign that led to the discovery of two new speculative leaks.
It would be interesting to investigate the effectiveness of these improvements, which are largely orthogonal to coverage guidance, within our framework.

Hofmann et al.~\cite{hofmann2023speculation} adapted the contract framework and contract-aware fuzzing to consider the effects of exceptions. 
They started with a simple model, refining it each time a new, unpredicted leak is found. 
This process allowed them to systematically document transient behaviors and discover three speculative vulnerabilities related to exceptions.
Our work currently does not consider exceptions, but it could be adapted in a similar manner.

Oleksenko et al.~\cite{oleksenko2025enter} present a fuzzing tool to find leaks across hardware isolation boundaries, such as between virtual machines or from the kernel to user space. 
Their framework uses ``actors" to represent different security domains and templates to create tests that involve domain transitions. 

\subsection{Pre-Silicon Hardware Fuzzing}
Existing work in pre-silicon fuzzing can further be classified into targeting functional bugs or side-channel vulnerabilities.

\paragraph{Functional Bugs}
DifuzzRTL~\cite{DBLP:conf/sp/HurSKBKL21}, which our work is based upon, aims to find functional bugs, i.e., cases where the processor's behavior deviates from a ``golden model'' that defines the expected ISA-level behavior. 
A witness to such a violation is a single executable, on which the processor and the ISA model deviate.
In contrast, contract violations are witnessed by pairs of executables that are contract indistinguishable, but lead to attacker-distinguishable executions.
We adapted DifuzzRTL accordingly.
Similarly, microarchitectural coverage notions for detecting functional bugs consider aspects of a single circuit's execution, such as multiplexer-toggle coverage~\cite{DBLP:conf/iccad/LaeuferKKBS18} or control register coverage~\cite{DBLP:conf/sp/HurSKBKL21}.
In contrast, our notion of self-composition deviation coverage measures differences between two concurrently executing instances of the same design.

Cascade~\cite{DBLP:conf/uss/SoltCR24} identifies two weaknesses of existing hardware fuzzing approaches: 1. The share of generated instructions that are actually executed during fuzzing is low. 
2. A large share of the instructions that do get executed are non-randomized initial and final code sequences.
Based on these insights they propose a two-stage program generation scheme that allows to generate long-running programs that do execute all of the generated instructions.
With growing program size, the overhead due to fixed initial and final code sequences shrinks.
Cascade achieves greater coverage than DifuzzRTL with no coverage guidance and uncovers 37 new bugs.
Incorporating and evaluating a similar program-generation approach in our framework remains future work.

\paragraph{Side-Channel Vulnerabilities}
SIGFuzz~\cite{rajapaksha2023sigfuzz} searches for instructions that influence the execution time of adjacent instructions by comparing the timing behavior of two programs that differ in a single instruction.
Like our work SIGFuzz builds upon DifuzzRTL.
They discover five side channels in Rocket and BOOM reusing the instrumentation and coverage metrics from DifuzzRTL.
They discover operand-dependent timing of division instructions in Rocket and BOOM, but miss a similar dependence of multiplication instructions.
They also note that ``If a load instruction is followed by another load instruction, the earlier load gets delayed."
Our minimized contract violations for BOOM often have two loads, but further investigation is needed to find out whether this is due to the same phenomenon.

SpecDoctor, by Hur et al. \cite{hur2022specdoctor}, is another variant of DifuzzRTL modified to find transient execution vulnerabilities. 
To this end, they statically detect components that can influence an instruction's timing behavior and focus the coverage on these components.
They perform experiments on BOOM and NutShell-Argo, and find novel Spectre attacks for both.

Both of the approaches above use the DifuzzRTL base and experiment on BOOM, yet their attacker models differ as the first approach allows timing-dependent behavior of a single instruction, while the second only searches for whole vulnerabilities that are able to transmit data.
We, in contrast, consider every data-dependent behavior that affects the overall timing of the program as a threat.

\section{Conclusions}

We introduce coverage-guided hardware-software contract fuzzing.
Coverage guidance leads to more efficient discovery of leaks and is enabled by targeting pre-silicon open-source processors.

At the core of our contribution is a security-aware coverage metric, self-composition deviation, which tracks divergences in the microarchitectural state between two processor instances, allowing our fuzzer to prioritize test cases that are more likely to cause contract violations.

Our experiments on the RISC-V Rocket and BOOM cores offer validation for this new approach. We first showed that coverage guidance does increase cumulative coverage, indicating a more efficient exploration of the processor's state space.  %
Our findings also establish a link between greater coverage and the discovery of contract violations in the complex out-of-order BOOM core.\looseness=-1

\section*{Use of Generative AI}

In the preparation of this document, an AI language model was utilized as a writing aid. The authors provided original drafts and source materials to the AI to assist with rephrasing and improving the clarity of the text. All AI-generated outputs were reviewed and verified by the authors.

\section*{Acknowledgements}

This project has received funding from the European Research Council (\url{https://erc.europa.eu/}) under the European Union's Horizon 2020 research and innovation programme (grant agreement no. 101020415).

\bibliographystyle{IEEEtranS}
\bibliography{references}

% Generated by IEEEtranS.bst, version: 1.14 (2015/08/26)
\begin{thebibliography}{10}
\providecommand{\url}[1]{#1}
\csname url@samestyle\endcsname
\providecommand{\newblock}{\relax}
\providecommand{\bibinfo}[2]{#2}
\providecommand{\BIBentrySTDinterwordspacing}{\spaceskip=0pt\relax}
\providecommand{\BIBentryALTinterwordstretchfactor}{4}
\providecommand{\BIBentryALTinterwordspacing}{\spaceskip=\fontdimen2\font plus
\BIBentryALTinterwordstretchfactor\fontdimen3\font minus
  \fontdimen4\font\relax}
\providecommand{\BIBforeignlanguage}[2]{{%
\expandafter\ifx\csname l@#1\endcsname\relax
\typeout{** WARNING: IEEEtranS.bst: No hyphenation pattern has been}%
\typeout{** loaded for the language `#1'. Using the pattern for}%
\typeout{** the default language instead.}%
\else
\language=\csname l@#1\endcsname
\fi
#2}}
\providecommand{\BIBdecl}{\relax}
\BIBdecl

\bibitem{AlmeidaBBDE16}
J.~B. Almeida, M.~Barbosa, G.~Barthe, F.~Dupressoir, and M.~Emmi, ``Verifying
  constant-time implementations,'' in \emph{Proceedings of the 25th {USENIX}
  Security Symposium}, ser. {USENIX} Security.\hskip 1em plus 0.5em minus
  0.4em\relax {USENIX} Association, 2016.

\bibitem{chipyard}
A.~Amid, D.~Biancolin, A.~Gonzalez, D.~Grubb, S.~Karandikar, H.~Liew,
  A.~Magyar, H.~Mao, A.~Ou, N.~Pemberton, P.~Rigge, C.~Schmidt, J.~Wright,
  J.~Zhao, Y.~S. Shao, K.~Asanovi\'{c}, and B.~Nikoli\'{c}, ``Chipyard:
  Integrated design, simulation, and implementation framework for custom
  {SoCs},'' \emph{IEEE Micro}, vol.~40, no.~4, pp. 10--21, 2020.

\bibitem{armstrong2019isa}
A.~Armstrong, T.~Bauereiss, B.~Campbell, A.~Reid, K.~E. Gray, R.~M. Norton,
  P.~Mundkur, M.~Wassell, J.~French, C.~Pulte \emph{et~al.}, ``{ISA} semantics
  for {ARMv8-a}, {RISC-V}, and {CHERI-MIPS},'' \emph{Proceedings of the ACM on
  Programming Languages}, vol.~3, no. POPL, pp. 1--31, 2019.

\bibitem{DBLP:journals/pacmpl/OlmosBBGL25}
\BIBentryALTinterwordspacing
S.~Arranz{-}Olmos, G.~Barthe, L.~Blatter, B.~Gr{\'{e}}goire, and V.~Laporte,
  ``Preservation of speculative constant-time by compilation,'' \emph{Proc.
  {ACM} Program. Lang.}, vol.~9, no. {POPL}, pp. 1293--1325, 2025. [Online].
  Available: \url{https://doi.org/10.1145/3704880}
\BIBentrySTDinterwordspacing

\bibitem{bernstein_aes_attack}
D.~J. Bernstein, ``Cache-timing attacks on {AES},'' 2005,
  \url{https://cr.yp.to/antiforgery/cachetiming-20050414.pdf}.

\bibitem{DBLP:journals/iacr/BloemGGHMP22}
\BIBentryALTinterwordspacing
R.~Bloem, B.~Gigerl, M.~Gourjon, V.~Hadzic, S.~Mangard, and R.~Primas, ``Power
  contracts: Provably complete power leakage models for processors,''
  \emph{{IACR} Cryptol. ePrint Arch.}, p. 565, 2022. [Online]. Available:
  \url{https://eprint.iacr.org/2022/565}
\BIBentrySTDinterwordspacing

\bibitem{buiras2021micro}
\BIBentryALTinterwordspacing
P.~Buiras, H.~Nemati, A.~Lindner, and R.~Guanciale, ``Validation of
  side-channel models via observation refinement,'' in \emph{{MICRO} '21: 54th
  Annual {IEEE/ACM} International Symposium on Microarchitecture, Virtual
  Event, Greece, October 18-22, 2021}.\hskip 1em plus 0.5em minus 0.4em\relax
  {ACM}, 2021, pp. 578--591. [Online]. Available:
  \url{https://doi.org/10.1145/3466752.3480130}
\BIBentrySTDinterwordspacing

\bibitem{DBLP:journals/tissec/DoychevKMR15}
\BIBentryALTinterwordspacing
G.~Doychev, B.~K{\"{o}}pf, L.~Mauborgne, and J.~Reineke, ``{CacheAudit}: {A}
  tool for the static analysis of cache side channels,'' \emph{{ACM} Trans.
  Inf. Syst. Secur.}, vol.~18, no.~1, pp. 4:1--4:32, 2015. [Online]. Available:
  \url{https://doi.org/10.1145/2756550}
\BIBentrySTDinterwordspacing

\bibitem{DBLP:conf/nanoarch/FuAJG21}
\BIBentryALTinterwordspacing
W.~Fu, O.~Arias, Y.~Jin, and X.~Guo, ``Fuzzing hardware: Faith or reality? :
  Invited paper,'' in \emph{{IEEE/ACM} International Symposium on Nanoscale
  Architectures, {NANOARCH} 2021, AB, Canada, November 8-10, 2021}.\hskip 1em
  plus 0.5em minus 0.4em\relax {IEEE}, 2021, pp. 1--6. [Online]. Available:
  \url{https://doi.org/10.1109/NANOARCH53687.2021.9642252}
\BIBentrySTDinterwordspacing

\bibitem{DBLP:conf/ccs/GarciaBY16}
\BIBentryALTinterwordspacing
C.~P. Garc{\'{\i}}a, B.~B. Brumley, and Y.~Yarom, ``{Make} sure {DSA} signing
  exponentiations really are constant-time,'' in \emph{Proceedings of the 2016
  {ACM} {SIGSAC} Conference on Computer and Communications Security, Vienna,
  Austria, October 24-28, 2016}, E.~R. Weippl, S.~Katzenbeisser, C.~Kruegel,
  A.~C. Myers, and S.~Halevi, Eds.\hskip 1em plus 0.5em minus 0.4em\relax
  {ACM}, 2016, pp. 1639--1650. [Online]. Available:
  \url{https://doi.org/10.1145/2976749.2978420}
\BIBentrySTDinterwordspacing

\bibitem{DBLP:conf/sp/GuarnieriKMRS20}
\BIBentryALTinterwordspacing
M.~Guarnieri, B.~K{\"{o}}pf, J.~F. Morales, J.~Reineke, and A.~S{\'{a}}nchez,
  ``Spectector: Principled detection of speculative information flows,'' in
  \emph{2020 {IEEE} Symposium on Security and Privacy, {SP} 2020, San
  Francisco, CA, USA, May 18-21, 2020}.\hskip 1em plus 0.5em minus 0.4em\relax
  {IEEE}, 2020, pp. 1--19. [Online]. Available:
  \url{https://doi.org/10.1109/SP40000.2020.00011}
\BIBentrySTDinterwordspacing

\bibitem{DBLP:conf/sp/GuarnieriKRV21}
\BIBentryALTinterwordspacing
M.~Guarnieri, B.~K{\"{o}}pf, J.~Reineke, and P.~Vila, ``Hardware-software
  contracts for secure speculation,'' in \emph{42nd {IEEE} Symposium on
  Security and Privacy, {SP} 2021, San Francisco, CA, USA, 24-27 May
  2021}.\hskip 1em plus 0.5em minus 0.4em\relax {IEEE}, 2021, pp. 1868--1883.
  [Online]. Available: \url{https://doi.org/10.1109/SP40001.2021.00036}
\BIBentrySTDinterwordspacing

\bibitem{hofmann2023speculation}
J.~Hofmann, E.~Vannacci, C.~Fournet, B.~K{\"o}pf, and O.~Oleksenko,
  ``Speculation at fault: Modeling and testing microarchitectural leakage of
  {CPU} exceptions,'' in \emph{32nd USENIX Security Symposium (USENIX Security
  23)}, 2023, pp. 7143--7160.

\bibitem{Hsiao21}
Y.~Hsiao, D.~P. Mulligan, N.~Nikoleris, G.~Petri, and C.~Trippel,
  ``Synthesizing formal models of hardware from {RTL} for efficient
  verification of memory model implementations,'' in \emph{MICRO}, 2021.

\bibitem{DBLP:conf/micro/HsiaoNKMPFT24}
\BIBentryALTinterwordspacing
Y.~Hsiao, N.~Nikoleris, A.~Khyzha, D.~P. Mulligan, G.~Petri, C.~W. Fletcher,
  and C.~Trippel, ``{RTL2M{\(\mu\)}PATH}: Multi-{\(\mu\)}path synthesis with
  applications to hardware security verification,'' in \emph{57th {IEEE/ACM}
  International Symposium on Microarchitecture, {MICRO} 2024, Austin, TX, USA,
  November 2-6, 2024}.\hskip 1em plus 0.5em minus 0.4em\relax {IEEE}, 2024, pp.
  507--524. [Online]. Available:
  \url{https://doi.org/10.1109/MICRO61859.2024.00045}
\BIBentrySTDinterwordspacing

\bibitem{hur2022specdoctor}
J.~Hur, S.~Song, S.~Kim, and B.~Lee, ``{SpecDoctor}: Differential fuzz testing
  to find transient execution vulnerabilities,'' in \emph{Proceedings of the
  2022 ACM SIGSAC Conference on Computer and Communications Security}, 2022,
  pp. 1473--1487.

\bibitem{hur2021difuzzrtl}
J.~Hur, S.~Song, D.~Kwon, E.~Baek, J.~Kim, and B.~Lee, ``{DifuzzRTL}:
  Differential fuzz testing to find {CPU} bugs,'' in \emph{2021 IEEE Symposium
  on Security and Privacy (SP)}.\hskip 1em plus 0.5em minus 0.4em\relax IEEE,
  2021, pp. 1286--1303.

\bibitem{DBLP:conf/sp/HurSKBKL21}
\BIBentryALTinterwordspacing
------, ``{DifuzzRTL}: Differential fuzz testing to find {CPU} bugs,'' in
  \emph{42nd {IEEE} Symposium on Security and Privacy, {SP} 2021, San
  Francisco, CA, USA, 24-27 May 2021}.\hskip 1em plus 0.5em minus 0.4em\relax
  {IEEE}, 2021, pp. 1286--1303. [Online]. Available:
  \url{https://doi.org/10.1109/SP40001.2021.00103}
\BIBentrySTDinterwordspacing

\bibitem{kocher2020spectre}
P.~Kocher, J.~Horn, A.~Fogh, D.~Genkin, D.~Gruss, W.~Haas, M.~Hamburg, M.~Lipp,
  S.~Mangard, T.~Prescher \emph{et~al.}, ``Spectre attacks: Exploiting
  speculative execution,'' \emph{Communications of the ACM}, vol.~63, no.~7,
  pp. 93--101, 2020.

\bibitem{DBLP:conf/iccad/LaeuferKKBS18}
\BIBentryALTinterwordspacing
K.~Laeufer, J.~Koenig, D.~Kim, J.~Bachrach, and K.~Sen, ``{RFUZZ:}
  coverage-directed fuzz testing of {RTL} on {FPGAs},'' in \emph{Proceedings of
  the International Conference on Computer-Aided Design, {ICCAD} 2018, San
  Diego, CA, USA, November 05-08, 2018}, I.~Bahar, Ed.\hskip 1em plus 0.5em
  minus 0.4em\relax {ACM}, 2018, p.~28. [Online]. Available:
  \url{https://doi.org/10.1145/3240765.3240842}
\BIBentrySTDinterwordspacing

\bibitem{Li:EECS-2016-9}
\BIBentryALTinterwordspacing
P.~S. Li, A.~M. Izraelevitz, and J.~Bachrach, ``Specification for the {FIRRTL}
  language,'' Tech. Rep. UCB/EECS-2016-9, Feb 2016. [Online]. Available:
  \url{http://www2.eecs.berkeley.edu/Pubs/TechRpts/2016/EECS-2016-9.html}
\BIBentrySTDinterwordspacing

\bibitem{Mohr24}
G.~Mohr, M.~Guarnieri, and J.~Reineke, ``Synthesizing hardware-software leakage
  contracts for {RISC-V} open-source processors,'' in \emph{Design, Automation
  and Test in Europe Conference and Exhibition (DATE), 2024}.\hskip 1em plus
  0.5em minus 0.4em\relax IEEE, Mar 2024.

\bibitem{DBLP:conf/icisc/MolnarPSW05}
\BIBentryALTinterwordspacing
D.~Molnar, M.~Piotrowski, D.~Schultz, and D.~A. Wagner, ``The program counter
  security model: Automatic detection and removal of control-flow side channel
  attacks,'' in \emph{Information Security and Cryptology - {ICISC} 2005, 8th
  International Conference, Seoul, Korea, December 1-2, 2005, Revised Selected
  Papers}, ser. Lecture Notes in Computer Science, D.~Won and S.~Kim, Eds.,
  vol. 3935.\hskip 1em plus 0.5em minus 0.4em\relax Springer, 2005, pp.
  156--168. [Online]. Available: \url{https://doi.org/10.1007/11734727\_14}
\BIBentrySTDinterwordspacing

\bibitem{Nemati2020a}
H.~Nemati \emph{et~al.}, ``Validation of abstract side-channel models for
  computer architectures,'' in \emph{{CAV}}, 2020.

\bibitem{riscv-tests}
\BIBentryALTinterwordspacing
T.~Newsome \emph{et~al.}, ``{riscv-tests},'' 2025, accessed: August 26, 2025.
  [Online]. Available: \url{https://github.com/riscv-software-src/riscv-tests}
\BIBentrySTDinterwordspacing

\bibitem{oleksenko2022revizor}
O.~Oleksenko, C.~Fetzer, B.~K{\"o}pf, and M.~Silberstein, ``Revizor: Testing
  black-box cpus against speculation contracts,'' in \emph{Proceedings of the
  27th ACM International Conference on Architectural Support for Programming
  Languages and Operating Systems}, 2022, pp. 226--239.

\bibitem{oleksenko2023hide}
O.~Oleksenko, M.~Guarnieri, B.~K{\"o}pf, and M.~Silberstein, ``Hide and seek
  with spectres: Efficient discovery of speculative information leaks with
  random testing,'' in \emph{2023 IEEE Symposium on Security and Privacy
  (SP)}.\hskip 1em plus 0.5em minus 0.4em\relax IEEE, 2023, pp. 1737--1752.

\bibitem{oleksenko2025enter}
O.~Oleksenko, F.~Solt, C.~Fournet, J.~Hofmann, B.~K{\"o}pf, and S.~Volos,
  ``Enter, exit, page fault, leak: Testing isolation boundaries for
  microarchitectural leaks,'' \emph{arXiv preprint arXiv:2507.06039}, 2025.

\bibitem{rajapaksha2023sigfuzz}
C.~Rajapaksha, L.~Delshadtehrani, M.~Egele, and A.~Joshi, ``Sigfuzz: A
  framework for discovering microarchitectural timing side channels,'' in
  \emph{2023 Design, Automation \& Test in Europe Conference \& Exhibition
  (DATE)}.\hskip 1em plus 0.5em minus 0.4em\relax IEEE, 2023, pp. 1--6.

\bibitem{DBLP:conf/fmcad/Reid16}
\BIBentryALTinterwordspacing
A.~Reid, ``Trustworthy specifications of {ARM}{\textregistered} {v8-A} and
  {v8-M} system level architecture,'' in \emph{2016 Formal Methods in
  Computer-Aided Design, {FMCAD} 2016, Mountain View, CA, USA, October 3-6,
  2016}, R.~Piskac and M.~Talupur, Eds.\hskip 1em plus 0.5em minus 0.4em\relax
  {IEEE}, 2016, pp. 161--168. [Online]. Available:
  \url{https://doi.org/10.1109/FMCAD.2016.7886675}
\BIBentrySTDinterwordspacing

\bibitem{DBLP:conf/uss/SoltCR24}
\BIBentryALTinterwordspacing
F.~Solt, K.~Ceesay{-}Seitz, and K.~Razavi, ``Cascade: {CPU} fuzzing via
  intricate program generation,'' in \emph{33rd {USENIX} Security Symposium,
  {USENIX} Security 2024, Philadelphia, PA, USA, August 14-16, 2024},
  D.~Balzarotti and W.~Xu, Eds.\hskip 1em plus 0.5em minus 0.4em\relax {USENIX}
  Association, 2024. [Online]. Available:
  \url{https://www.usenix.org/conference/usenixsecurity24/presentation/solt}
\BIBentrySTDinterwordspacing

\bibitem{DBLP:conf/asplos/TanYBM025}
\BIBentryALTinterwordspacing
Q.~Tan, Y.~Yang, T.~Bourgeat, S.~Malik, and M.~Yan, ``{RTL} verification for
  secure speculation using contract shadow logic,'' in \emph{Proceedings of the
  30th {ACM} International Conference on Architectural Support for Programming
  Languages and Operating Systems, Volume 1, {ASPLOS} 2025, Rotterdam, The
  Netherlands, 30 March 2025 - 3 April 2025}, L.~Eeckhout, G.~Smaragdakis,
  K.~Liang, A.~Sampson, M.~A. Kim, and C.~J. Rossbach, Eds.\hskip 1em plus
  0.5em minus 0.4em\relax {ACM}, 2025, pp. 970--986. [Online]. Available:
  \url{https://doi.org/10.1145/3669940.3707243}
\BIBentrySTDinterwordspacing

\bibitem{wang2023specification}
Z.~Wang, G.~Mohr, K.~von Gleissenthall, J.~Reineke, and M.~Guarnieri,
  ``Specification and verification of side-channel security for open-source
  processors via leakage contracts,'' in \emph{Proceedings of the 2023 ACM
  SIGSAC Conference on Computer and Communications Security}, 2023, pp.
  2128--2142.

\bibitem{wang2025}
------, ``Synthesis of sound and precise leakage contracts for open-source
  risc-v processors,'' in \emph{Proceedings of the 2025 ACM SIGSAC Conference
  on Computer and Communications Security}, 2025.

\bibitem{afl}
\BIBentryALTinterwordspacing
M.~Zalewski \emph{et~al.}, ``{AFL},'' 2025, accessed: August 29, 2025.
  [Online]. Available: \url{https://github.com/google/AFL}
\BIBentrySTDinterwordspacing

\end{thebibliography}

\end{document}